

\magnification=\magstep1
\openup 1\jot
\def\xb{{\bf x}}
\font\secbf=cmbx12
\font\subsecbf=cmbx10

\outer\def\beginsec#1\par{\vskip0pt plus.3\vsize\penalty-250
 \vskip0pt plus-.3\vsize\bigskip\vskip\parskip
 \message{#1}\leftline{\secbf#1}\nobreak\smallskip\noindent}

\outer\def\beginsubsec#1\par{\vskip0pt plus.1\vsize\penalty-100
 \vskip0pt plus-.1\vsize\medskip\vskip\parskip
 \message{#1}\leftline{\subsecbf#1}\nobreak\smallskip\noindent}

\rightline{LA-UR-94-2231}

\centerline{{\bf THE LEAST ACTION METHOD, CDM AND $\Omega$}}
\vskip 0.5 truein
\centerline{ A.M. Dunn\footnote{$^ 1$}{Email: amd@mail.ast.cam.ac.uk}}
\centerline{Institute of Astronomy}
\centerline{University of Cambridge, Madingley Road}
\centerline{Cambridge, CB3 0HA, UK}
\vskip 0.5 truein
\centerline{ R. Laflamme\footnote{$^ 2$}{Email: laf@t6-serv.lanl.gov}}
\centerline{Theoretical Astrophysics, T-6, MSB288}
\centerline{Los Alamos National Laboratory}
\centerline{Los Alamos, NM87545, USA}
\vskip 1 truein
\centerline{\bf Abstract}
\bigskip
Peebles has suggested an interesting method to trace back in time
positions of galaxies called the least action method.  This method
applied on the Local Group galaxies seems to indicate that we live in
an $\Omega\approx 0.1$ Universe.  We have studied a CDM N-body
simulation with $\Omega=0.2$ and $H=50 kms^{-1}/Mpc$ and compare
trajectories traced back from the Least Action Principle and the
center of mass of the particle forming CDM halos.  We show that the
agreement between these set of trajectories is at best qualitative.
We also show that the line of sight peculiar velocities are
underestimated.  This discrepancy is due to orphans, CDM particles
which do not end up in halos.  By varying the density parameter
$\Omega$ in the least action principle we show that using this method
we would underestimate the density of the Universe by a factor of 4-5.

\vfill\eject

\beginsec 1. Introduction

The density parameter $\Omega$ is one the most important parameter
characterising our Universe. There are many methods to measure $\Omega$,
but a factor of 10 uncertainty in its value remains. Recently Peebles suggested
that
it may be possible to estimate its value in the Local Neighborhood, by tracing
Local Group galaxies back in time.

Peebles (1989,1990, 1994) used the principle of least action to find
complete trajectories for Local Group galaxies.
The idea is to assume that galaxies growing out of small density perturbations
in the early universe will have negligible peculiar velocities with respect
to the   Hubble flow. This is a reasonable assumption as we know that the
microwave background has very small anisotropies.
Using zero initial peculiar velocities as one boundary condition and the
present
positions of the galaxies as the other, trial orbits are iteratively varied
so as to minimize the action. The method has been  criticised since the
galaxies are treated as point particles throughout their history, even
though the size of the galaxies must be comparable to their separation
at early times. However, the least action principle leaves the final
velocities of the galaxies unconstrained,
and its ability to reproduce the observed radial velocities
remains a  powerful  test of the validity of the trajectories. For the
Local  Group galaxies, Peebles has obtained remarkable agreement between the
observed radial velocities and those calculated from the least action
principle.  Obtaining reliable trajectories for nearby galaxies might
shed light into the origin of their angular momentum (Dunn $\&$ Laflamme,
1993).

Although the Least Action Method provides a powerful tool for investigating
galaxy orbits, its predictions are only as good as the assumptions they stand
on.
These are, that a) galaxies initially have negligible peculiar velocity with
respect to the Hubble flow, b) galaxies can be represented as point particles
throughout their history, c) mergers have little effect on a galaxies motion,
and
d) light traces mass. One way to test the validity of some of these assumptions
is to apply the Least Action Method to a numerical simulation of the universe.
In the simulation we have complete information about the particle trajectories,
which we compare with the predictions made by the Least Action Method.
In this letter we use a cold dark matter (CDM) simulation. Although CDM may not
be able to reproduce all the observable features of our universe, it does at
least represent a {\it possible} universe in which all the particles are
governed
by Hamiltonian dynamics.

In the first section we give details of the simulations
we have used and comment on the groups that we have studied. Secondly
we compare the trajectories obtained from the CDM simulation and
the Least Action Method and compare their `line of sight' velocities.
Finally we comment on the origin of the discrepancy.

\beginsec 2. The Least Action Method.

Peebles' use of the least action method (LAM) selects a set of classical
trajectories for a group of galaxies (point masses) which are interacting
through gravity, against
the background of an expanding universe model. This method differs from the
usual application of the least action principle in that boundary conditions
are applied to the beginning and end of each trajectory. The trajectories
are constrained such that
$$
\delta {\bf x}_i = 0 \ \ {\rm at} \ \ t = t_0, \hskip 1cm
a^2 d{\bf x}_i /dt \rightarrow 0 \ \ {\rm at} \ \ a\rightarrow 0
\eqno(2.1)
$$
where $a$ is the scale factor of the universe and ${\bf x}_i(a)$ is the
trajectory of the $i$th galaxy in comoving coordinates.
That is, the galaxies are fixed at their present positions at the present
epoch,
 and their peculiar velocities vanish as we approach the Big Bang.
Trial trajectories for the set of galaxies are adjusted in order to find
a stationary point in the action.

In this paper a matter dominated universe with no cosmological constant is
assumed, thus
$$
H dt = { a^{1/2} da \over F^{1/2}},
\eqno(2.2)
$$
where $F=(\Omega + (1-\Omega)a)$, $H$ is the present Hubble constant and
$\Omega$ is the density parameter.
Following Peebles (1990), the action for particles moving in such a universe
is,
$$
S= \int_0^{t_o} \Big [ \sum {m_i a^2 \over 2}
                \big ( {d{\bf x}_i \over dt} \big )^2
           + {G\over a} \sum_{i\neq j}{ m_im_j \over |{\bf x}_i - {\bf x}_j|}
           + {2\over 3} \pi G \rho_b a^2 \sum m_i{\bf x}_i^2 \Big ]
\eqno(2.3)
$$
from which we can deduce the equation of motion,
$$
a^{1/2}{d\over da} a^{3/2} {d \xb_i \over da}
 + {(1-\Omega)a^2 \over 2F}{d\xb_i\over da}
  = {\Omega\over 2F} [ \xb_i + {R^3_0\over M_T}
  \sum_j {m_j(\xb_j-\xb_i)\over |\xb_j-\xb_i|^3} ].
\eqno(2.4)
$$
Here $R_0$ is the radius of a sphere which would enclose a homogeneous
distribution of the total mass $M_T$ of the group of galaxies considered,
$R_0^3 \equiv M_T ({4 \over 3} \pi \rho^0_b)^{-1}$.  Note that this equation is
slightly different from the one used by Peebles as we do not assume a
flat ($k=0$) universe .

It is very hard to have exact analytic solutions for the coupled
system of equations (2.4).  However Peebles succeeded in obtaining approximate
solutions using trial functions of the form
$$
\xb_i(a) = \xb^o_i + \sum_n {\bf C}^n_i f_n(a)
\eqno(2.5)
$$
where $\xb^o_i$ are the present positions of the galaxies and the $f_n$
are linearly independent functions chosen to satisfy the boundary
conditions (2.1).   In this paper we take  $f_n = a^n (1-a)$ for $n=0,
\ldots ,4$. The classical solutions are obtained by introducing ${\bf
x}_i(a)$ in the action and iteratively modifying the coefficient  ${\bf
C}^n_i$ to obtain a stationary action.  As Peebles did, we verify that
the least action solutions are good approximations to real solutions by
evolving the classical equations of motion starting with the initial
positions and velocities  derived from the least action solutions at
$z=60$.

\beginsec 3. CDM simulation.

In order to understand the limits of the Least Action Method, it is important
to compare it with some other method.  We used a CDM, N-body
simulation of  Kauffmann and White (1992).
It is a PPPM simulation with 262 144 particle, representing
an $\Omega=0.2$ universe. Scaled to
$H=50 km\ s^{-1}\ Mpc^{-1}$, it encompasses a size of 100 Mpc  with particles
of mass $5.2\times 10^{10}\ M_\odot$.

We have studied a few groups containing 10 or so galaxy halos. The halos
are determined by a friend of a friends algorithm. They were chosen
in order to match the conditions in the Local Group; two dominant galaxies
with peculiar velocities towards each other, a mass ratio of roughly
4:3 and somewhat isolated from high density mass concentrations.
 Due to the limitations of dynamic range in the simulation,
 we could not find any such halos with a separation
of 0.7 Mpc but had to go to approximately 2 Mpc. These halos also
had masses approximately 5-10 times greater than M31 and the MW.
 In addition to the two central galaxies, the galaxies around them up to a
distance of 20 Mpc were selected to form a group.  We thus intend to study the
effect
of the
spatial distribution of the halo, the influence of the nearby galaxies on their
dynamics and also the effect of particles not linked to any halo (orphans).

We have selected 9 groups which had 2 halos of the order of 200 particles
within
2 Mpc.  We investigated roughly 10 galaxies around these two center halos.
All the groups had similar behavior and we therefore for brevity present the
results
of only one of them here.

\beginsec 4. Comparison.

Once we have identified the galactic halos we can use the least action method
described in section 2 to trace them back in time.  We can also trace back the
particles making the halo in the final step of the CDM simulation. Figure 1.
shows the 3 projections of the halos trajectories. We can see that there is a
very rough agreement between the LAM trajectories and the CDM ones.

Assuming that these are in rough agreement we can compare the line of sight
velocities of these  two trajectories.  It is seen from figure 2 that with the
same parameter $\Omega$ and $H$, the LAM would overestimate this velocity. By
dividing $\Omega$ by a factor of 4 we we would obtain reasonable line of sight
velocity. It is this problem that we address now.

First let's see the effect of modifying $\Omega$ in the LAM. By varying
$\Omega$
in the LAM we can change the line of sight velocities.  Here there are two
factors to take into account. Changing $\Omega$ will change the time elapsed
since the Big Bang, increasing  $\Omega$ decrease the elapsed time and thus
increased the velocity. The second factor is that with a larger $\Omega$ the
radius $R_0$ of eq. (2.4), the radius of a sphere which would enclose a
homogeneous distribution of
the total mass,  is smaller.  Thus we do not have to go as far to gather the
mass to make the halos, thus decreasing the velocity.  These two factors
conspire
against each other but a simple calculation for a 2 body system show that the
first one wins. Thus as shown in figure 2 as $\Omega$ decreases the velocity
decreases.

We must now answer the question of why, for the same $\Omega$ and $H$
we have the LAM line of sight velocities being larger than the CDM
one.  One incorrect justification would be to think that the radius
$R_o$ to gather the halo's mass, should be increased in proportion to
the fraction of CDM particles which do not end up in halos.  However
this is incorrect as can be shown by comparing the trajectories from
the CDM and LAP simulations.  The correct answer lies in the fact that
in the CDM model there are orphans, i.e.\ CDM particles which are not
linked to halos.  In the early times the background of orphans is
roughly homogeneously distributed and cancel the force due to the
particles which will eventually make the halos. From the LAP point of
view, this will in fact reduce the force or the effective mass of the
halos and thus, in order to end up at their known final positions,
they will have to start at a closer distance than the LAM has given
us.  This has the effect of reducing the velocities, as the CDM
simulation shows.

As mentioned earlier, we can also investigate the effect of the
spatial distribution of the halos.  Consider equation (2.4), all the
terms are linear (in $x^i$) except the last one on the right. We have
compared the contribution of this term, which we call by abuse of
language, the inhomogeneous component of force, when we do the sum
over particles in different ways. In the LAM we assume that the halos
interact as point sources, that is, the important part of the force
only acts between the centers of mass of each halo,\
$$
{\bf F}^{1}_a = \sum_{b} { {\bf x}_b -{\bf x}_a \over
                           | {\bf x}_b -{\bf x}_a |^3}
\eqno (4.1)
$$
where $a$ is the target galactic halo  and the sum over $b$ is over
the center of mass of the nearby halos.

The second approach is obtained by summing over all the particles in
each of the halos rather than just their center of mass.  We have also
divided by the number of particles of the target halo ($N_a$) to get the force
on its center of mass.  This will give an estimate of the effect of
the higher multiple moment of the halos.
$$
{\bf F}^{2}_a = {1\over N_a}
                   \sum_{a_i}\sum_{b_j} { {\bf x}_{b_j} -{\bf x}_{a_i} \over
                           | {\bf x}_{b_j} -{\bf x}_{a_i} |^3}
\eqno (4.2)
$$
here the sum over $a_i$ is over all particles of the target halo and
the one over ${b_j}$ is over all particles of the halo $b$ and then
over all halos in our sample.  This will essentially sum over
everything except the orphans.

The third quantity is
$$
{\bf F}^{3}_a = {1\over N_a}
                    \sum_{a_i}\sum_{j} { {\bf x}_{j} -{\bf x}_{a_i} \over
                           | {\bf x}_{j} -{\bf x}_{a_i} |^3}
\eqno (4.3)
$$
where the sum $a_i$ is over all particles of the target halo and the
sum ${j}$ is over all particles within 20Mpc of the CM of the target
group at $r=9.98465$ i.e., the last frame of the CDM model.  This
corresponds to the true force on the halo. We have modified the
distance of 20 Mpc to a shorter distance and without significant
change in the results (for the force on the last frame).

We have plotted the result in Figure 3 where the magnitude
of the different `forces' and the angle between the first two and
${\bf F}^{3}_a$ are shown.  From this figure we can see that at early times
the force ${\bf F}^{1}_a$ is overestimated by a factor of roughly 2, which
is not unexpected, since galaxies make poor approximations to point particles
at early times. We can also see that the force ${\bf F}^{2}_a$,
which includes higher multipoles of halos, is not a very
good approximation since there are serious discrepancies and scatter between
the direction of this vector and the true force ${\bf F}^{3}_a$.
 We must therefore reject the suggestion
of Branchini and   Carlberg (1994), that the discrepancy between
the CDM and LAM line of sight velocities might be due to neglecting the
shape of the CDM halos.

We should also point out that another possible problem for
the LAM is the existence of mergers.  In one of our groups there was a
significant merger and for this halo the force was not very well
represented by the one at its center of mass.  For a merger to have
an important effect it must be the result of roughly equally massive
halos which come in rather different directions.  A detailed study of
mergers in
CDM model is needed to quantitatively know if this is a potentially serious
problem for the LAM.

\beginsec 5. Conclusion.

In this letter we have shown that Peebles' Least Action Method
underestimates the value of $\Omega$ for a CDM universe.  The main discrepancy
is due to neglecting the effect of orphans, CDM particles which have are not
members of
any halos.  They are scattered uniformly in the early stage of the universe
and therefore reduce the force on the particles which will eventually form
halos.  Thus the proto-halos must start at a shorter distance than what is
expected in Peebles original suggestion.  This is equivalent to failure of
one of the key assumptions of the Least Action Method, that is, that light
traces mass (at least at kiloparsec to megaparsec scales).
Of course, it is quite possible that this assumption
{\it does} in fact hold for the universe we live in.
However, since there is little observational or theoretical reason to suppose
this, we must call in to doubt
previously published results based on the LAM. We conclude that
the dynamics of the Local Group and a careful examination of
its line of sight velocities do  not exclude a closed Universe.

\noindent{\bf Acknowledgments}.  We would like to thank B. Bromley,
D. Lynden-Bell, S. White, M. Warren
and W.H. Zurek for useful comments. We would also like to thank the
NASA HPCC program for support.

\beginsec 5. References

\noindent
Branchini, E. \&  Carlberg, R.G., Testing the Least Action Principle in and
$\Omega_o=1$ Universe, SISSA preprint 56-94-A.

\noindent
Dunn, A.M. \& Laflamme, R., 1993, {\it M.N.R.A.S.}, {\bf 264}, 865.

\noindent
Kauffman, G \& White, S.D.M., 1992, {\it M.N.R.A.S.}, {\bf 258}, 511.

\noindent
Peebles, P.J.E., 1989, {\it Astrophys.J.}, {\bf 344},L53.

\noindent
Peebles, P.J.E., 1990, {\it Astrophys.J.}, {\bf 362}, 1.

\noindent
Peebles, P.J.E., 1994, Orbits of nearby galaxies,
{\it Astrophys.J.}, to appear.

\vfill\eject
\proclaim Figure captions.

\vskip 0.25 truein \openup -1\jot \noindent
{\narrower\smallskip\noindent Figure 1.  Projection of the CDM and least
action trajectories for galaxies in the chosen group. The broken line
are the trajectories from the CDM simulation by following the center
of mass of the particles forming a halo in the last frame.  The plain
line are the LAM trajectories.  We can see that the agreement is at
best qualitative. (Units in Mpc).  \smallskip} \openup 1\jot \vskip 0.25 truein

\vskip 0.25 truein \openup -1\jot \noindent
{\narrower\smallskip\noindent Figure 2. Least Action Method line of
sight velocities (with respect to one target galaxy of the chosen
group) as a function of the CDM line of sight velocity.  The best fit
corresponds to an adjusted density parameter $\Omega\approx 0.05$ a
factor of 4-5 higher than the CDM simulation parameter.  \smallskip}
\openup 1\jot
\vskip 0.25 truein

\vskip 0.25 truein \openup -1\jot \noindent
{\narrower\smallskip\noindent Figure 3. Plot of the inhomogeneous
part of the `force' (eq. [4.1-3]) $F^1_a$ and $F^2_a$
with respect to the true force $F^3_a$ on a typical halo.  We see that
the force on a halo is not well approximated by the force due to other
halos,  orphans (CDM particles not bound to halos) have an important
contribution.   \smallskip} \openup 1\jot
\vskip 0.25 truein

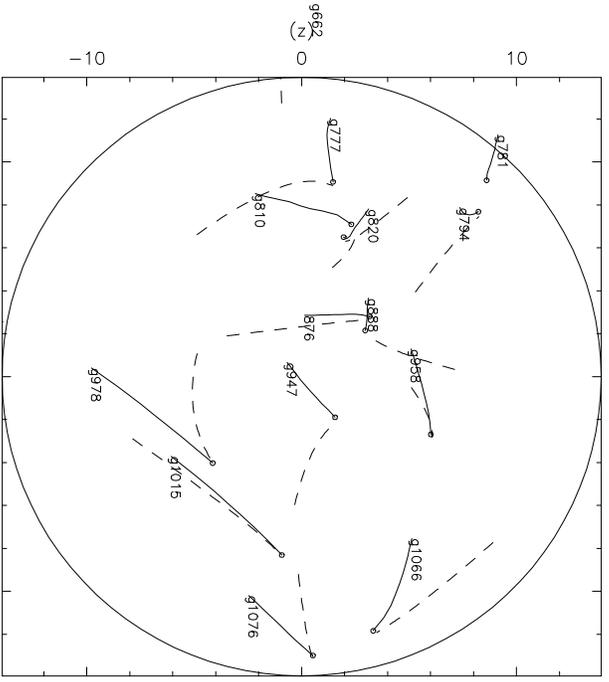
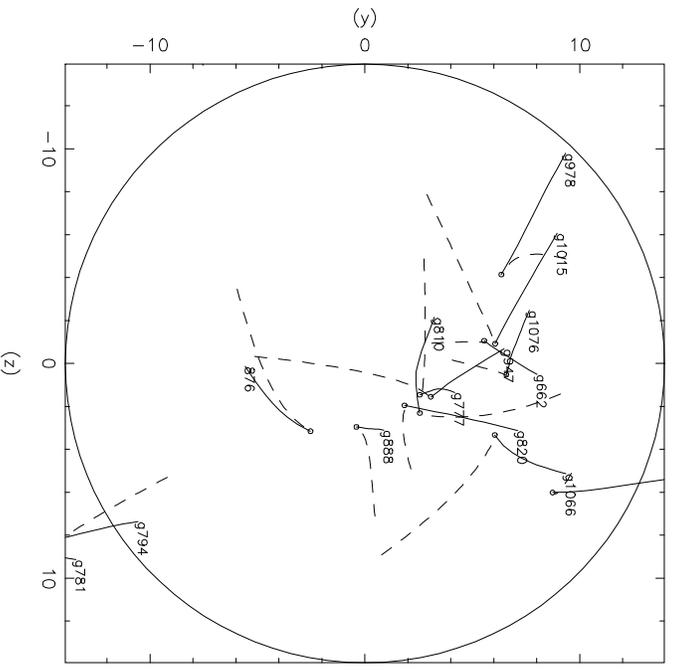
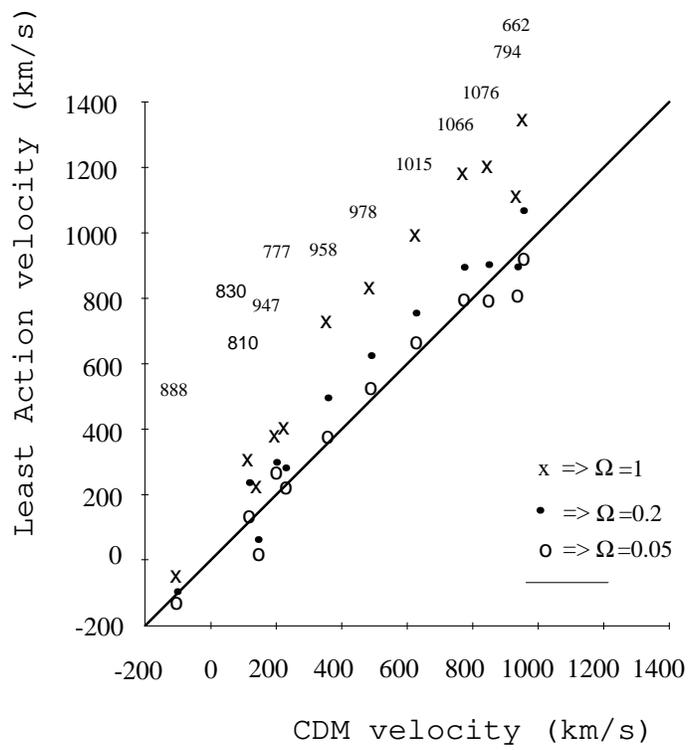
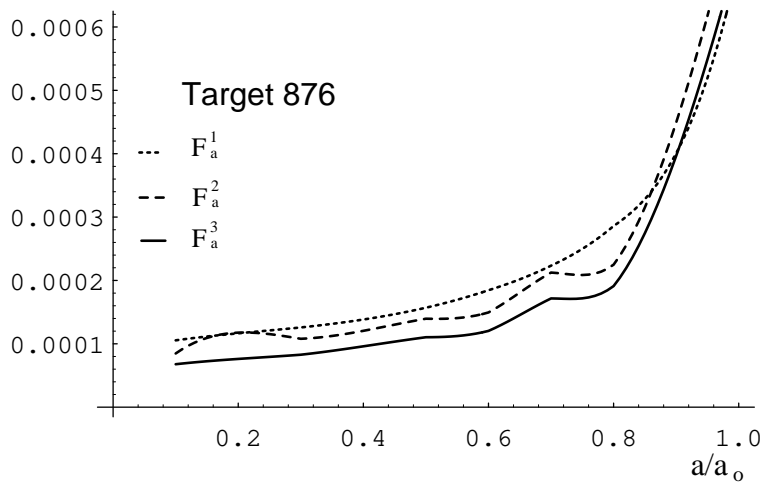
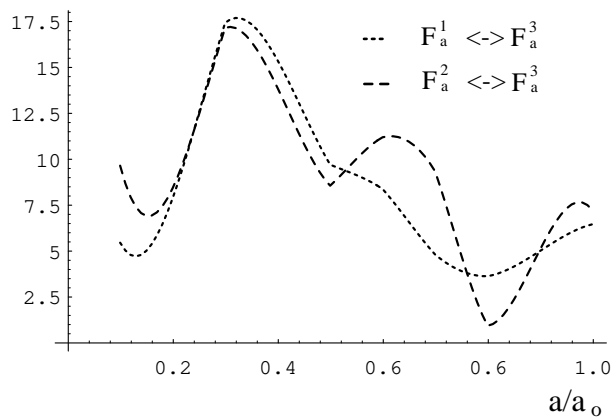
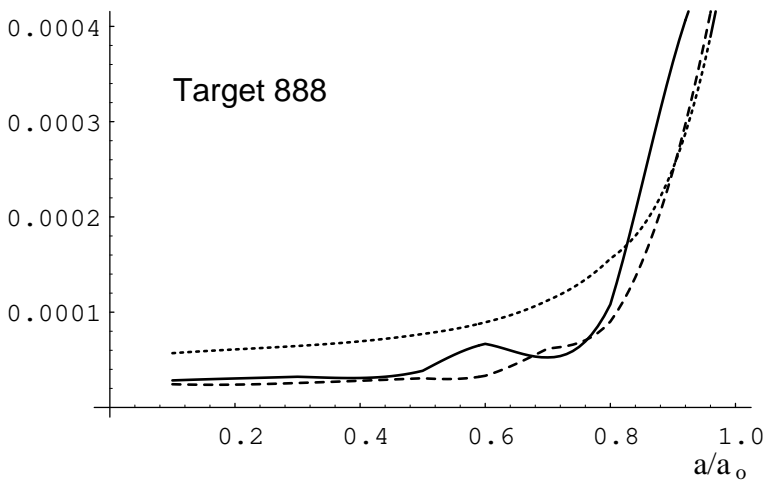
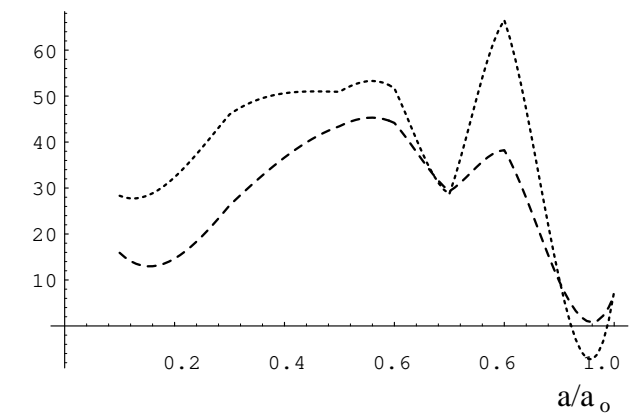
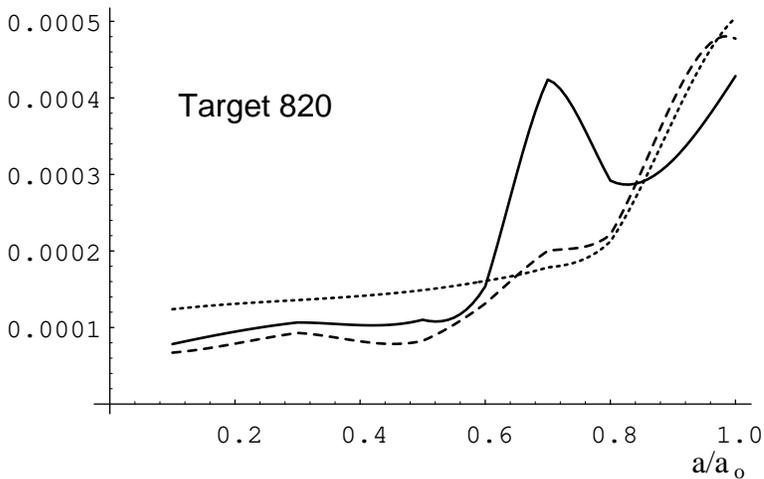
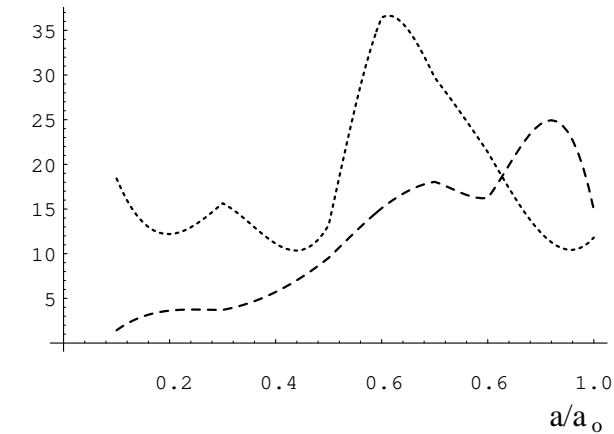
\bye